\documentclass[a4,12pt]{article}

\usepackage[T1]{fontenc}
\usepackage[utf8]{inputenc}

\usepackage{authblk}
\usepackage{graphicx}
\usepackage{subcaption}
\usepackage{amssymb}
\usepackage{multirow}
\usepackage{booktabs}
\usepackage[normalem]{ulem}
\usepackage{cite}
\usepackage{amsmath}
\usepackage{xcolor}

\title{$T_{cc}$ and Hidden Charm Tetraquarks $1^+$ and $0^+$ in QCD sum rules and Heavy-Quark Spin Symmetry}

\author{Semra Sarı Ferah\footnote{ssferah@metu.edu.tr (corresponding author)}}
\author{Altug Ozpineci\footnote{ozpineci@metu.edu.tr} }
\affil{Physics Department, Middle East Technical University, 06800, Ankara, Türkiye}
\date{\today}
\begin{document}

\maketitle
\begin{abstract}
In this work, $T_{cc}$ double charm tetraquark and $1^{+}$ and $0^{+}$ hidden charm tetraquarks are studied within the QCD sum rules framework. 
In the heavy-quark limit, the $1^+$ and $0^+$  tetraquarks are degenerate and the difference in their masses is an $1/m_c$ effect. It is shown that, although the
uncertainty in the mass predictions of each of these hadrons is ${\cal O}(100\;\mathrm{MeV})$, when the mass differences are studied, 
the uncertainty is reduced to less than ${\cal O}(10\;\mathrm{MeV})$. Hence, it is concluded that if one of the exotic
hadrons is observed and its mass is used as input, the masses of the other hadrons can be determined with a precision of 
${\mathcal O}(10\; \mathrm{MeV})$
using the QCD sum rules method.

\end{abstract}
\section{Introduction}\label{Sec:Introduction}

Recent years have been an exciting period for hadronic spectroscopy due to the discovery of many hadrons that cannot be easily described as conventional mesons or baryons. 
Among these newly discovered states, there are some states that are definitely exotic, such as the double open charm meson $T_{cc}$ \cite{LHCb:2021vvq}, and some whose internal structure is not so easily understood, such as the $\chi_{c1}(3872)$ state (formerly known as X(3872)) \cite{Belle:2003nnu}

Understanding the structure of these particles has been a challenge. 
In the case of X(3872), although it has been more than a decade since its discovery, its structure is still an active area of research (see, e.g. \cite{Ji:2022uie, Baru:2015nea,Zanetti:2011ju, Colangelo:2022awx} and references therein).
Being non-perturbative objects, their study requires the use of non-perturbative methods.  
Effective theories, heavy-quark symmetry, and QCD sum rules are among the methods used to shed light on the structure of newly discovered exotic states \cite{nieves2012heavy, hidalgo2013light, hidalgo2013x, yamaguchi2015heavy, hosaka2016exotic, mutuk2018x, aliev2024qcd}.

Quarks whose masses are heavier than the scale of QCD, $\Lambda_{QCD}$, are considered as heavy quarks. 
In the limit that their masses are sent to infinity, a new symmetry of QCD, namely the "Heavy-Quark Symmetry", emerges. 
In this limit, the spin of the heavy quark decouples from the dynamics and states that differ only in the spin of the heavy quark become degenerate \cite{Neubert:1993mb}. 
This allows one to obtain relations between various exotic hadrons \cite{nieves2012heavy,Jia:2023hvc, mutuk2018x}. 
Finite-mass effects lead to splitting between masses.

A non-perturbative method that has been widely used to study properties of hadrons is the QCD sum rules method, first developed by M. A. Shifman, A. I. Vainshtein, and V. I. Zakharov for conventional mesons in 1979 \cite{shifman1979qcd} and generalized to conventional baryons by B. L. Ioffe in 1981 \cite{ioffe1981calculation} . 
In this method, one studies a suitable chosen correlation function and relates the properties of the hadrons to properties of the vacuum through a handful of condensates. QCD sum rules has been extensively used to study the
properties of exotic hadrons (see e.g. \cite{navarra2007qcd,mutuk2018x,Agaev:2021vur}).

The QCD sum rules method preserves the heavy-quark symmetry that appears in the large quark mass limit. 
In \cite{hidalgo2013light}, the degeneracy between the heavy-quark spin symmetry partners of X(3872) has been demonstrated starting from a correlation function, which was later analyzed in \cite{mutuk2018x} to obtain the predictions of the masses from the correlation function. 
In \cite{hidalgo2013light,mutuk2018x}, the states analyzed have light degrees of freedom in the total spin state $s=1$, whereas the heavy degrees of freedom can be in the $s=0$ or $s=1$ state.
In this work, the states in which the light degrees of freedom are in the $s=0$ state are analyzed. 
The heavy quarks can be in a total spin $s=1$ state (leading to a particle with $J^{P}=1^{+}$ ) or in a total spin $s=0$ state (leading to a particle with $J^{P}=0^{+}$).
In the heavy-quark limit, these two particles will degenerate.
Note that in the tetraquark $T_{cc}$, the light degrees of freedom can also be in a total spin $s=0$ state 
similar to the above two states. Hence, in this work, the double charm $T_{cc}$ and hidden charm $1^+$ and $0^+$ tetraquarks will be analyzed within the QCD sum rules.

In section \ref{sumrules}, QCD sum rules analysis will be presented, followed by a discussion on the implications of the heavy-quark symmetry for the correlation functions in section \ref{hqt}. 
Finally, in section \ref{sec:Numerical Analysis and Conlusions} we present our numerical results and conclusions.

\section{QCD Sum Rules}

\label{sumrules}
To obtain the spectrum of particles, the following correlation functions will be studied:
\begin{align} \label{eq:correlation Function of 1+}
\Pi_{\mu \nu}^{(T)}\left(p^2\right)&= i \int d^4 x e^{i p x}\left\langle \Omega\left|\mathcal{T}\left\{J^{(T)}_{\mu}(x)
{J^{(T)}_{\nu}}^\dagger\left(0\right)\right\}\right|\Omega \right\rangle,
\\
\label{eq:correlation Function of 0+}
\Pi\left(p^2\right)&= i \int d^4 x e^{i p x}\left\langle \Omega\left|\mathcal{T}\left\{J(x)
{J}^\dagger \left(0\right)\right\}\right|\Omega \right\rangle,
\end{align}
where, $J^{(T)}_\mu$  and $J$ are the interpolating currents that can create the axial vector ($T_{cc}$) and scalar tetraquarks from the vacuum, repectively, $p$ is the four momentum of the hadron,  $\vert\Omega \rangle$ is the physical non-perturbative hadronic vacuum and $\mathcal{T}$ is the time ordering operator.

The interpolating currents that will be used in this study are chosen as:
\begin{align} \label{eq:Interpolating current of tcc}
  J^T_{\mu}\left(x\right) &= \left[\left(c^{aT} C \gamma_\mu c^b\right)\left(x\right)\left( { \bar{u}}^{c } \gamma_5 C \bar{d }^{dT}\right)\left(x\right)\right] \varepsilon^{a b e} \varepsilon^{c d e},  
\\ 
  J_{\mu}\left(x\right) &= \left[\left(\bar{c}^{a} \gamma_\mu c^c\right)\left(x\right)\left( { \bar{u}}^{b } \gamma_5 {d }^d\right)\left(x\right)\right] \varepsilon^{a b e} \varepsilon^{c d e},  
\\ 
\label{eq:Interpolating current of xcc0}
  J\left(x\right) &= \left[\left(\bar{c}^{a} \gamma_5 c^{c}\right)\left(x\right)\left( { \bar{u}}^{b } \gamma_5 {d }^d\right)\left(x\right)\right] \varepsilon^{a b e} \varepsilon^{c d e},  
\end{align}
where, a, b, c, d, and e are color indices and $C$ is the charge conjugation matrix. The light quarks created by these operators are all in $s=0$ state. Note that $J^T_\mu$ is an isoscalar operator, whereas $J_\mu$ and $J$ are isovector operators.

In QCD sum rules, the correlation function is calculated in two different kinematical regions, and then the two representations are matched \cite{cohen1995qcd}. For $p^2>0$, the correlation function can be expressed in terms of the properties of hadronic degrees of freedom, the so-called hadronic (phenomenological) representation; and for $p^2 \ll 0$, it can be calculated in terms of the QCD parameters, the so-called OPE representation.

In the kinematical region $p^2>0$, by inserting a complete set of hadronic states between the interpolating currents in the correlation functions, they can be expressed as: 
\begin{align} \label{eq:Phen part of xcc1}
    \Pi^{(T)Had}_{\mu \nu}\left(p^2\right)&=\frac{|{\tilde \lambda^{(T)}}_{0}|^2 p_\mu p_\nu}{p^2-\tilde m_{0}^2}+\frac{|\lambda_{1}^{(T)}|^2 \left(-g_{\mu\nu}+\frac{p_\mu p_\nu}{p^2}\right)}{p^2-m_{(T)1^{+}}^2}\nonumber\\ &\equiv \tilde \Pi^{(T)}_0 p_\mu p_\nu + \Pi^{(T)}_1 \left(-g_{\mu\nu} + \frac{p_\mu p_\nu}{p^2}\right),\\ 
 \label{eq:Phen part of xcc0 2}  \Pi^{Had}\left(p^2\right)&=\frac{|\lambda_{0}|^2 }{p^2-{m_{0^{+}}^2} } \equiv \Pi_0.
\end{align}
Note that since the current $J^{(T)}_\mu$ couples to both particles with $J^{P}=0^{+}$ and $J^{P}=1^{+}$ (both for the hidden charm and $T_{cc}$),
there are two contributions to the correlation functions $\Pi_{\mu\nu}^{(T)Had}$, and for a given $J^P$ quantum number, 
only the contribution of the lowest mass hadrons are shown explicitly and the contributions from higher states and the continuum are not shown.

In order to obtain the OPE representation of the correlation function, heavy (light) quark propagators are needed. 
Their explicit expression are \cite{reinders1985hadron, cohen1995qcd, colangelo2001qcd}:
\begin{align}\label{eq:full propagator}
S_{full}^{a b, q}(x)= &  \frac{i\not\!x}{2 \pi^2 x^4} \delta^{a b}-\frac{\langle q \bar{q}\rangle}{12}
\delta^{a b} -\frac{x^2}{192} m_0^2\langle q \bar{q}\rangle \delta^{a b} 
\nonumber\\\nonumber\\ & -i g_s \int_0^1 d u\left[\frac{\not\!x}{16 \pi^2 x^2} G^{a b}_{\alpha \beta}(u x) \sigma^{\alpha \beta}-u x_\mu G^{a b}_{\alpha \beta}(u x) \gamma^\nu \frac{i}{4 \pi^2 x^2}\right]+\cdots , \\\nonumber\\ 
 S_{full}^{a b, c}(x)  &=\left[\frac{m_c^2}{4 \pi^2} \frac{K_1\left(m_c \sqrt{-x^2}\right)}{\sqrt{-x^2}}\right] -i\left[ \frac{m_c^2}{4 \pi^2 x^2} K_2\left(m_c \sqrt{-x^2}\right)\right]\not\! x \nonumber\\\nonumber\\ &-i g_s \int \frac{d^4 k}{(2 \pi)^4} e^{-i k x} \int_0^1 d v\left[\frac{\not\!k+m_c}{\left(m_c^2-k^2\right)^2} G^{a b}_{\alpha \beta}(v x) \sigma^{\alpha \beta}\right. \nonumber\\\nonumber\\ 
& \left.+\frac{1}{m_c^2-k^2} v x_\alpha G^{a b}_{\alpha \beta} \gamma^\beta\right] +\cdots , 
\end{align}
where the light quark masses are set to zero and $K_1$ and $K_2$ are the modified Bessel functions of the second kind.

The two expressions for the correlations functions can be matched using their spectral representation and for each of the function $\Pi_0$ and $\Pi_1$, the sum rules are obtained as:
\begin{equation} \label{eq:next definition OPE=Phen}
\lambda_h^2 e^{-m_h^2 / M^2}= \int_0^{s_0} d s \rho_h^{\mathrm{OPE}}(s) e^{-s / M^2}.
\end{equation}
After the Borel transformation, which is carried out to eliminate the unknown polynomials in the spectral representation, continuum subtraction is done using quark-hadron duality (see, e.g. \cite{shifman2001quark} for more details). 
In Eq. \ref{eq:next definition OPE=Phen}, $h=0$,  $h=1$ or $T$ for the hidden charm scalar, hidden charm axial vector and $T_{cc}$ tetraquarks.
The explicit forms of the spectral densities $\rho_h^{OPE}(s)$ are given in the appendix.

Once the spectral densities are obtained, the mass of the hadron $h$ can be calculated using the following relation:
\begin{align} \label{eq:Mass Calculation} 
  m_h^2=\frac{\int_0^{s_0} ds s \rho_h^{\text {OPE }}(s)e^{-{s/M^2}}}{\int_0^{s_0} ds \rho_h^{\text {OPE }}(s)e^{-{s/M^2}}}.
\end{align}
where $s_0$ is the continuum threshold and $M^2$ is the Borel parameter.

\section{Heavy-Quark Limit}
\label{hqt}
In the heavy-quark limit, the interpolating currents for the hidden charm states take the form
\begin{align} \label{eq:HQInterpolating current of xcc1}
  J_{\mu}^{v}\left(x\right) &= \left[\left(\bar h_{\bar c}^a\gamma_\mu h_c^c\right)\left(x\right)\left( { \bar{q_1}}^{b } \gamma_5 {q_2 }^d\right)\left(x\right)\right] \varepsilon^{a b e} \varepsilon^{c d e}, \\  
 \label{eq:HQInterpolating current of xcc0}
  J^{0,v}\left(x\right) &= \left[\left(\bar h_{\bar c}^a \gamma_5 h_c^c \right)\left(x\right)\left( { \bar{q_1}}^{b } \gamma_5 {q_2 }^d\right)\left(x\right)\right] \varepsilon^{a b e} \varepsilon^{c d e},  
\end{align}
where $h_c$($\bar h_{\bar c}$) are annihilation operators for the the heavy (anti) charm quark with velocity $v$.
Note that in the heavy quark limit, $v^\mu J_\mu^v=0$, hence $J_\mu$ can only create (axial) vector particles from the vacuum, and hence there are no contributions to the correlation function $\Pi_{\mu\nu}$ from scalars, i.e. $\tilde \Pi_0=0$ exactly.

In the heavy quark limit, the correlation functions formed by the above currents can be expressed as (see e.g. \cite{hidalgo2013light}) 
\begin{eqnarray}
    \Pi_{\mu\nu} &=& i \int e^{i p x} \langle 0 \vert {\cal T} J_\mu(x) J_\nu^\dagger(0)\vert 0 \rangle \equiv \mbox{Tr} \left(\frac{1+\not\!v}{2} \gamma_\mu \frac{1-\not\!v}{2} \gamma_\nu\right) {\cal R}(p^2)\nonumber \\ &=& 2 \left(g_{\mu\nu} - v_\mu v_\nu \right) {\cal R}(p^2),
    \\
    \Pi &=& i \int e^{i p x} \langle 0 \vert {\cal T} J(x) J^\dagger(0)\vert 0 \rangle \equiv \mbox{Tr}\left(\frac{1+\not\!v}{2} \gamma_5 \frac{1-\not\!v}{2} \gamma_5\right) {\cal R}(p^2) \nonumber \\ &=& 2 {\cal R}(p^2).
\end{eqnarray}
Note that since both currents have the same light quark structure, and the effects of the spin of the heavy quark is taken into account in the  traces in above equations, the function ${\cal R}(p^2)$ appearing in the correlation function in the above expression are the same functions.
Compared with the hadronic representation, Eqs. \ref{eq:Phen part of xcc1} and \ref{eq:Phen part of xcc0 2}, it is seen that  
\begin{eqnarray} 
\Pi_{1} &=& -2 R(p^2), \\
\Pi_{0} &=& 2 R(p^2),
\end{eqnarray}
i.e. the contributions of the axial vector particles to the interpolating current $\Pi_{\mu\nu}$ and the contributions of the scalars to $\Pi$ are equal to each other (upto an overall constant). 
In particular, as a function of $p^2$, the poles of both correlation functions are at the same position in the heavy-quark limit, proving the degeneracy of the $1^{+}$ and $0^{+}$ states.

To obtain the heavy quark limit expression of the correlation functions from the finite quark mass expression, it is enough to make the redefinitions \cite{neubert1992heavy} 
\begin{equation}
    \begin{split}
     & (2 m_c)T\equiv M^2,\\
     & (2 m_c) v_0\equiv s_0-(2 m_c)^2,\\
     & (2 m_c) \tilde{\Lambda}\equiv m_h-(2m_c)^2,
    \end{split}
\end{equation}
and take the limit $m_c \rightarrow \infty$, keeping the leading term in $1/m_c$. 

Note that heavy-quark symmetry does not relate the hidden charm states to the $T_{cc}$ state.

\section{Numerical Analysis and Conlusions}
\label{sec:Numerical Analysis and Conlusions}
QCD sum rules has many input parameters, including quark masses and quark/gluon condensates in the QCD vacuum. 
For calculations of physical quantities, numerical values of these parameters are needed, and the values used in this study are listed in Table \ref{tab:Input parameters used in calculations} \cite{particle2022review, shifman1979qcd, narison2010gluon,ParticleDataGroup:2024cfk}. 

\begin{table}[h!]
 \caption{Input parameters used in calculations}
     \label{tab:Input parameters used in calculations}
     \centering
     \begin{tabular}{|c|c|}
\hline \hline \textbf{Parameters} & \textbf{Values} \\
\hline \hline
\hline$m_u$ & 0 \\
\hline$m_d$ & 0\\
\hline$m_c$ & $1.27 \pm 0.02 \;\mathrm{GeV}$ \\
\hline$\langle\bar{u} u\rangle$ & $(-0.24 \pm 0.01)^3 \;\mathrm{GeV^3}$ \\
\hline$\langle\bar{d} d\rangle$ & $(-0.24 \pm 0.01)^3 \;\mathrm{GeV^3}$ \\
\hline$m_0^2$ & $(0.8 \pm 0.1)\;\mathrm{GeV^2}$ \\
\hline$\left\langle g_s^2 G^2\right\rangle$ & $4 \pi^2(0.012 \pm 0.004) \;\mathrm{GeV^4}$\\
\hline \hline
\end{tabular}
 \end{table}

In addition to these parameters, the sum rules expressions also contain two auxiliary parameters: the Borel parameter $M^2$ and the continuum threshold $s_0$.  
The Borel parameter $M^2$ is an arbitrary parameter, and the results should be independent of the value of this parameter. 
On the other hand, the continuum threshold $s_0$ parameterizes the energy above which excited states and the continuum start. 
To determine the mass of the hadrons using QCD sum rules, a suitable range of these parameters should be determined such that the predictions are independent of these parameters.

For the continuum threshold, the typical values used in the sum rules are determined by the relation:
\begin{align}\label{eq:s_0 interval}
(m_h+0.3\; \mathrm{GeV)^2} \le s_0\le (m_h+0.5\; \mathrm{GeV)^2},
\end{align}
which in our case corresponds to $s_0=(19\pm1)\;\mathrm{GeV^2}.$ 

\begin{figure}%
\hspace*{-15mm}
\begin{center}
\includegraphics{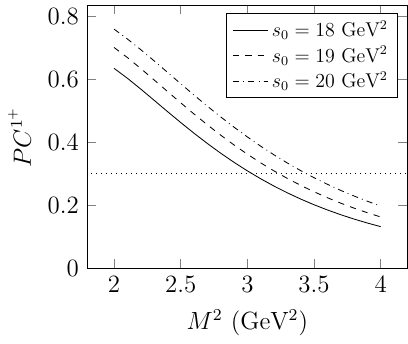}%
\includegraphics{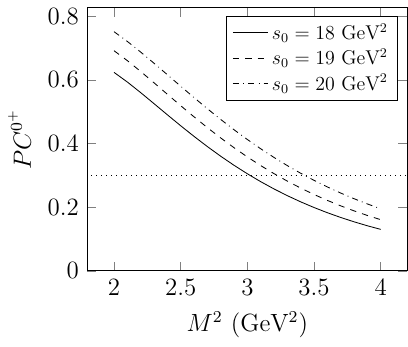}
\includegraphics{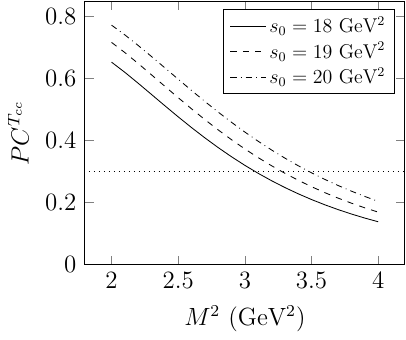}
\end{center}
\caption{The pole contribution to each correlation function plotted as a function of the Borel parameter $M^2$ evaluate at $s_0=18$, $19$ and $20$ GeV$^2$}
    \label{fig:PC}%
\end{figure}

For the working region of $M^2$, the following criteria are considered:
If $M^2$ is too small, the OPE expansion does not converge. 
Hence, the lower limit is obtained by requiring that perturbative contributions to the correlation functions are dominant. 
However, if $M^2$ is too large, the contributions of higher states and the continuum are not suppressed. 
Hence, requiring the pole contribution to be at least $30\%$ gives an upper bound of the allowed values of $M^2$. 
And finally, there should be a subregion within this region such that the predictions are independent of the value of $M^2$.
To determine the maximum $M^2$ value,  the pole contribution (PC) is defined as:
\begin{align} \label{eq:Pole Contribution}
PC^h=\frac{\int_{0}^{s_0} ds  \rho_h(s)e^{-{\frac{s}{M^2}}}}{\int_{0}^{\infty} ds \rho_h(s)e^{-{\frac{s}{M^2}}}},
\end{align}
where $h=T_{cc}$, $0^+$ or $1^+$ is analyzed. 
In Fig. \ref{fig:PC}, the pole contribution to the correlation functions for $1^+$, $0^+$ and $T_{cc}$ tetraquarks are shown as a function of $M^2$. 
As can be seen in both of these figures, $PC \gtrsim 30\%$ for $M^2<3.5\;\mathrm{GeV^2}$ if $s_0=18$ GeV$^2$ and for $M^2<3.0$ GeV$^2$ if $s_0=20$ GeV$^2$. Since the predictions obtained are practically independent of the value of $M^2$ for $3.0$ GeV$^2<M^2<3.5$ GeV$^2$, the upper limit
of the working region for the Borel parameter will be chosen as $M^2=3.5$ GeV$^2$.

\begin{figure}%
\hspace*{-15mm}
\begin{center}
\includegraphics{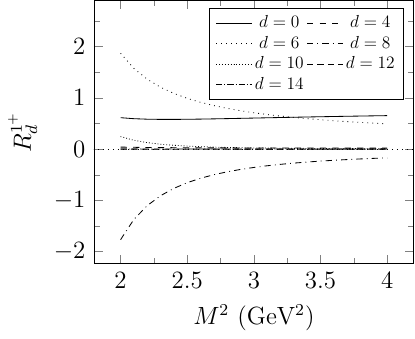}%
\includegraphics{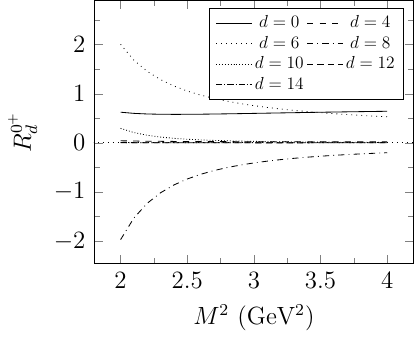}
\includegraphics{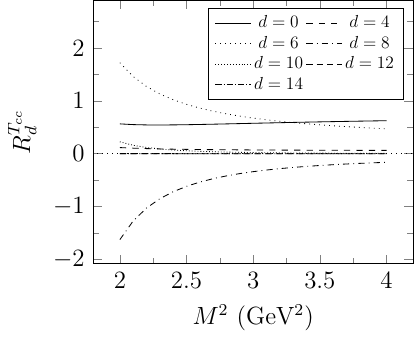}
\end{center}
   \caption{The convergence of the contribution of each dimension separately obtained as a function of $M^2$ at $s_0=19\;\mathrm{GeV^2}$ value for ${1^{+}}$, ${0^{+}}$ and $T_{cc}$ tetraquarks }%
     \label{fig:Convergence}%
\end{figure}

To determine the lower limit of the working region of the Borel parameter $M^2$, the convergence of the OPE is examined. 
The relative contribution of the operator of dimension $d$ is defined as:
\begin{align} \label{eq:Convergence} R^h_d\left(M^2\right)=\frac{\Pi_h^{d}[M^2,s_0]}{\Pi_h[M^2,s_0]},
\end{align}
where $\Pi^{d}_h(M^2,s_0)$ is the contribution of the $d$ dimensional condensate to the OPE.
The plots of $R^h_d$ are shown in Fig. \ref{fig:Convergence}.  
As can be seen from the figures, within the plotted range, the perturbative contribution is more than half of the total result. 
Although the operators $d=6$ and $d=8$  are comparable to the perturbative result, since they have opposite signs, they cancel each other out. Still, requiring that $\vert R^d_h\vert<1$ leads to the lower limit $M^2>2.5\;\mathrm{GeV^2}$.
Even higher dimensional operators are negligible and can be neglected.

\begin{figure}%
\hspace*{-15mm}
\begin{center}
\includegraphics{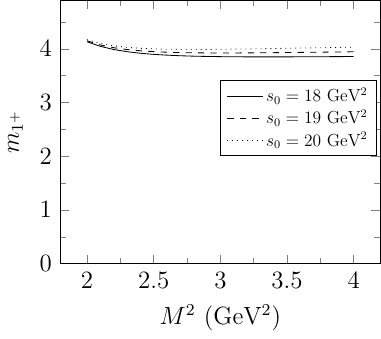}%
\includegraphics{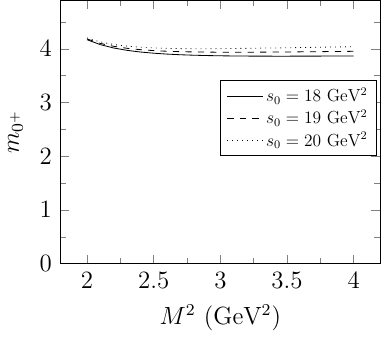}
\includegraphics{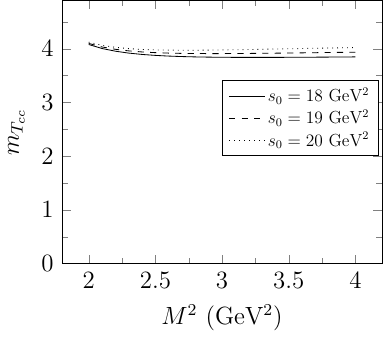}
\end{center}
\caption{$M^2$ dependence of masses of $1^+$, $0^+$ and $T_{cc}$ tetraquarks for $s_0=18$, $19$ and $20$ GeV$^2$}
     \label{fig:msqdependence}%
\end{figure}

Finally, in Fig. \ref{fig:msqdependence}, $M^2$  dependences of the calculated masses for the $T_{cc}$, $1^{+}$ and $0^{+}$ particles are shown for $s_0=18$, $19$ and $20$ $\mathrm{GeV^2}$. 
It can be easily noticed that for the chosen working regions the mass values are practically independent of the Borel parameters. 


\begin{figure}
  \hspace*{-15mm}
  \begin{center}
  \includegraphics{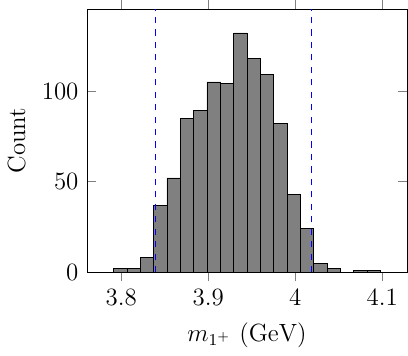}%
  \includegraphics{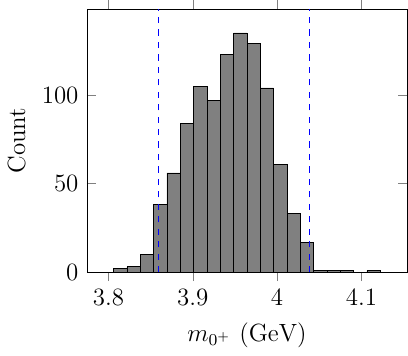}
  \includegraphics{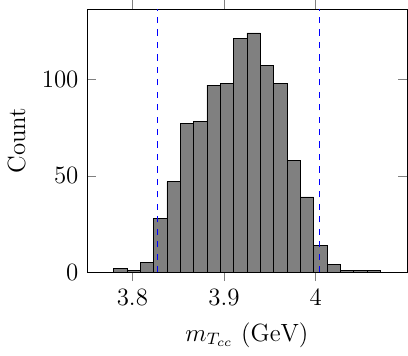}
  \end{center}
\caption
{Histograms for the predicted masses of the $1^{+}$, $0^{+}$ and  $T_{cc}$ tetraquarks.}
\label{fig:histograms}
\end{figure}

Figs. \ref{fig:PC}, \ref{fig:Convergence}, and \ref{fig:msqdependence}  are obtained for the central values of the parameters shown in Table \ref{tab:Input parameters used in calculations}. 
In order to obtain all uncertainties including those caused by uncertainties in the input parameters, the analysis proposed in \cite{Leinweber:1995fn} is followed. 
For this purpose, 1000 sets of random values are chosen for the parameters $m_c$, $\langle \bar u u \rangle=\langle \bar d d \rangle$, $m_0^2$, $\langle g_s^2 G^2 \rangle$ within the ranges shown in the table, and $M^2$ and $s_0$ are chosen within the determined working regions. 
The distribution of condensates and $m_c$ are assumed to be Gaussian with the mean equal to the central value and the standard deviation equal to half the uncertainty; therefore, more than $80\%$ of the values lie within the uncertainties shown in the table. 
The distribution of $M^2$ and $s_0$ is assumed to be flat within the determined range. 
The histograms of the mass values obtained for the states ${1^{+ }}$, ${0^{+ }}$, and ${T_{cc}}$ are shown in Fig. \ref{fig:histograms}. 
Using the mean and twice the standard deviation of the distributions, as the central value and uncertainty, the masses of the axial vector and scalar particles are predicted to be $m_{1^{+ }}=3.93 \pm 0.09 \;\mathrm{GeV} $ and $m_{0^{+ }}=3.95\pm0.09 \;\mathrm{GeV}$,
$m_{T_{cc }}=3.92\pm0.09 \;\mathrm{GeV}$. These regions are shown as vertical dashed lines in the histograms.

\begin{figure}
    \centering
    \includegraphics[width=0.5\linewidth]{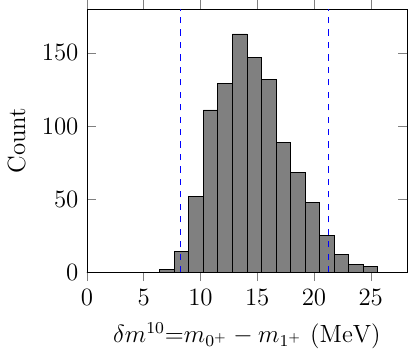}%
    \includegraphics[width=0.5\linewidth]{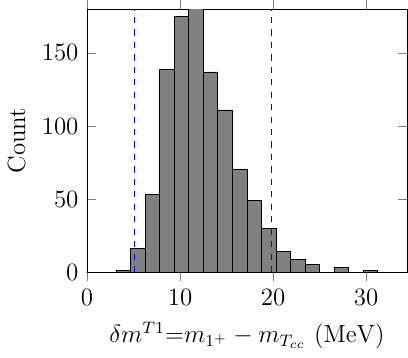}
    \includegraphics[width=0.5\linewidth]{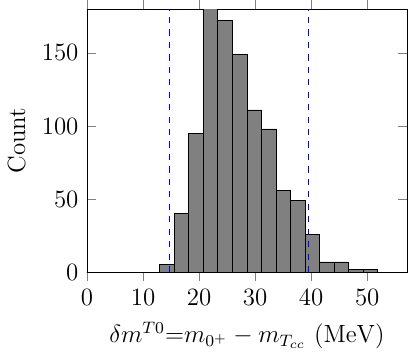}
    \caption{Histograms for the predicted mass differences.}
    \label{fig:massdifference}
\end{figure}

Furthermore, in Fig. \ref{fig:massdifference}, the histograms of the mass differences of these particles are depicted. 
The mass differences are predicted to be 
\begin{align}
 \delta m^{10}& \equiv m_{0^{+}}-m_{1^{+}}=14.8\pm 6.5\;\mathrm{MeV}, \nonumber \\
 \delta m^{T1} & \equiv m_{1^+} - m_{T_{cc}} =\nonumber 12.5 \pm 7.4\;\mathrm{MeV}, \\
 \delta m^{T0} & \equiv m_{0^+} - m_{T_{cc}} = 27.1 \pm 12.4 \;\mathrm{MeV}.
\end{align}

Note that the uncertainty in the mass difference $\delta m^{10}$ is  smaller than the uncertainties in the other two mass differences. This is due to the fact that $\delta m^{10}$ is protected by heavy-quark symmetry. Nevertheless, 
even though the mass differences $\delta m^{T1}$ and $\delta m^{T0}$ are not protected by heavy-quark
symmetry, the corresponding uncertainties are still ${\cal O}(10\;\mathrm{MeV})$. Using the experimentally measured value of the $T_{cc}$ mass $m_{T_{cc}}=3874.75 \pm 0.10\;\mathrm{MeV}$ \cite{ParticleDataGroup:2026pdg}, and the mass differences, the masses of the $1^+$ and $0^+$ hidden charm tetraquark states can be obtained as:
\begin{align}
    &m_{1^+} = m_{T_{cc}} + \delta m^{T1} = 3887.3\pm7.5\;\mathrm{MeV}, \nonumber \\
    &m_{0^+} = m_{T_{cc}} + \delta m^{T0} = 3901.9\pm 12.5 \;\mathrm{MeV}. 
\label{eq:predictedmasses}
\end{align}

Such a precise determination of the masses simplifies the identification of the particles by comparing the predicted masses with the experimentally
measured masses. 
Our predictions for the masses of the ${T_{cc}}$, ${1^{+ }}$, and ${0^{+ }}$ states, along with the  results from various experimental and theoretical works, are presented in Tables \ref{tab:Tcc}, \ref{tab:1Plus}, and, \ref{tab:0Plus} respectively.
In the tables, the row labeled ``This work'' represents the masses obtained using Eq. \ref{eq:Mass Calculation} directly whereas ``This work - using $m_{T_{cc}}$ as input''  shows the values presented in Eq. \ref{eq:predictedmasses}. 

\begin{table}[h!]
\centering
\caption{The measured and predicted mass of the $T^+_{cc}$ tetraquark }
\label{tab:Tcc}
\begin{tabular}{p{0.65\textwidth} p{0.3\textwidth}} 
\toprule 
Experimental and Theoretical Works & $M\left(\mathrm{MeV} / c^2\right)$ \\
\midrule 
LHCB \cite{collaboration2022study} & $3875.1 \pm 0.7 $ \\
\cite{agaev2022newly} & $3868 \pm 124$ \\
\cite{navarra2007qcd} & $4000 \pm 200$ \\
\cite{dias2011relation} & $3872.2 \pm 39.5$ \\
\cite{karliner2017discovery} & $3882 \pm 12$ \\
\cite{eichten2017heavy} & $3978$ \\
\cite{wang2018analysis} & $3900 \pm 90$ \\
\cite{braaten2021masses} & $3947 \pm 11$ \\
\cite{cheng2021double} & $3929.3$ \\
\bottomrule 
PDG \cite{ParticleDataGroup:2026pdg}  & $3874.75 \pm 0.10$  \\
\textbf{This work} & $\mathbf{3920 \pm 90}$ \\
\bottomrule
\end{tabular}
\end{table}
\begin{table}[h!]
\centering
\caption{The predictions of the mass of the $1^+$ states obtained from different experimental and theoretical works}
\label{tab:1Plus}
\resizebox{\textwidth}{!}{%
\begin{tabular}{p{0.55\textwidth} p{0.4\textwidth}}
\toprule
Experimental and Theoretical Works & $M\left(\mathrm{MeV} / c^2\right)$ \\
\midrule
BESIII \cite{ablikim2013observation}  & $3899 \pm 6 $ \\
Belle \cite{liu2013study} & $3895 \pm 8$  \\
CLEO-c \cite{xiao2013observation} & $3886\pm 5$ \\
BESIII \cite{ablikim2014observation}  & $3884 \pm 5 $ \\
BESIII \cite{ablikim2015confirmation}  & $3882 \pm 3 $ \\
D0 \cite{boos2019properties} & $3902.6^{+5.2}_{-5.0}$ \\
BESIII \cite{ablikim2020study} & $3893.0\pm 2.3 \pm 3.2$ \\
BESIII \cite{ablikim2026partial} & $3884.6\pm 0.7 \pm 3.3$ \\
\multirow{2}{*}{\cite{xiao2013observation}} & $3886\pm4\pm 2 $ \\ 
& $3904\pm9\pm 5$ \\
\cite{wang2014analysis} & $3910^{+110}_{-90} $ \\
\cite{aceti2014prediction} & $3878$ \\
\cite{wang2015analysis} & $3910^{+210}_{-170}$ \\
\cite{he2015z} & $3876 $ \\
\multirow{2}{*}{\cite{albaladejo2016zc}} & $3894 \pm 6 \pm 1$ \\
& $3886 \pm 4 \pm 1$ \\
\cite{agaev2016strong}& $3900 \pm 210 $ \\ 
\multirow{3}{*}{\cite{ikeda2016fate}} & $3709 \pm 94 $ \\
& $3748 \pm 76 $ \\
& $3686 \pm 56 $ \\ 
\multirow{3}{*}{\cite{pilloni2017amplitude}}  & $3893.2 ^{+5.5}_{-7.7} $ \\
& $3905 ^{+11}_{-9} $ \\
& $3900 ^{+140}_{-90} $ \\
\cite{ortega2019z} & $3871.74$ \\
\cite{chen2024precise} & $3880.7\pm1.7\pm 22.4$ \\ 
\midrule
PDG \cite{ParticleDataGroup:2024cfk}  & $3887.1 \pm 2.6$ \\
\textbf{This work-direct determination} & $\mathbf{3930 \pm 90}$ \\
\textbf{This work-using $m_{T_{cc}}$ as input}   & $\mathbf{3887.3 \pm 7.5}$  \\
\bottomrule
\end{tabular}%
}
\end{table}

\begin{table}[h!]
\centering
\caption{The predictions of the mass of the $\mathrm{0}^+$ states obtained from different experimental and theoretical works}
\label{tab:0Plus}
\resizebox{\textwidth}{!}{%
\begin{tabular}{p{0.6\textwidth} p{0.3\textwidth}}
\toprule
Experimental and Theoretical Works & $M\left(\mathrm{MeV} / c^2\right)$ \\
\midrule
Belle \cite{choi2005observation} & $3943 \pm 11 \pm 13$ \\
BaBar \cite{del2010evidence} & $3919.1_{-3.4}^{+3.8} \pm 2.0$ \\
Belle \cite{uehara2010observation} & $3915 \pm 3 \pm 2$ \\
BaBar \cite{lees2012study} & $3919.4 \pm 2.2 \pm 1.6$ \\
LHCb \cite{aaij2020amplitude} & $3923.8 \pm 1.5 \pm 0.4$ \\
\cite{li2024determination} & $3920.9 \pm 0.9$ \\
\cite{yang2010dynamical} & $3912.1$ \\
\multirow{2}{*}{\cite{ortega2018charmonium}} & $3890.3$ \\
& $3889.0$ \\
\multirow{2}{*}{\cite{duan2020possibility}} & $3904$ \\
& $3885$ \\
\cite{wang2020analysis} & $3950 \pm 90$ \\
\midrule
PDG \cite{ParticleDataGroup:2024cfk} & $3921.7 \pm 1.8$ \\
\textbf{This work-direct determination} & $\mathbf{3950 \pm 90}$ \\
\textbf{This work-using $m_{T_{cc}}$ as input} & $\mathbf{3901.9 \pm 12.5}$ \\
\bottomrule
\end{tabular}%
}
\end{table}

As can be seen in the tables \ref{tab:1Plus} and \ref{tab:0Plus},
calculating the differences and using one mass as experimental input drastically reduces the uncertainty in sum rules predictions. 

The hidden charm isovector state $T_{c\bar c1}(3900)$ has an observed mass of $m_{T_{c\bar c 1}}=3887.1\pm2.6\;\mathrm{MeV}$ \cite{ParticleDataGroup:2024cfk} which coincides exactly with the predicted mass given in Eq. \ref{eq:predictedmasses}.
 For the case of the scalar particle, among the observed particles, the closest is $X(3915)$ with an observed mass of $m_{X(3915)}=3922.1\pm1.8\;\mathrm{MeV}$ \cite{ParticleDataGroup:2024cfk}. To understand the discrepancy with the predicted value, note that the observed particle has $I=0$, whereas
the particle studied in this work has $I=1$. 
Within a sum rules analysis, the mass separation between the $I=0$ and $I=1$ particles of the same quark content is created by the annihilation diagrams. 
In general, for the study of masses directly, ignoring annihilation diagrams is justified but if one is interested in studying mass differences that are
${\cal O}(10\;\mathrm{MeV})$, they have to be included in the computations.
Annihilation diagrams are in general ignored in the sum rules analysis as they will be small.
Nevertheless, in \cite{Zhang:2024fxy}, an isovector scalar is predicted to exist, but a mass prediction is not presented.
In this work, we predict its mass to be $ m_{0^+} = 3901.9\pm 12.5 \;\mathrm{MeV}$.

In conclusion, sum rules does not allow one to determine the mass of a particle with a precision better than a few hundred $\mathrm{MeV}$.
However, if the mass differences are calculated, the uncertainty is ${\cal O}(10\;\mathrm{MeV})$ or less. If the mass difference is protected by a symmetry (such as the heavy-quark symmetry) it is even reduced to a few MeV.
Hence, using the mass of an observed exotic hadron as an input, the masses of other exotics can be calculated to a precision of just several MeV if they are heavy-quark symmetry partners.

In this work, the masses of doubly open charm $T_{cc}$ and hidden charm $1^{+}$ and $0^{+}$ tetraquarks are analyzed using QCD sum rules. Using the observed mass of $T_{cc}$ as input, the mass of the $1^+$ 
isovector axial tetraquark is predicted to be $m_{1^+} = 3887.3\pm7.5\;\mathrm{MeV}$, which is consistent with the
experimentally observed $T_{c\bar c1}(3900)$. Furthermore, the mass of the hidden charm isovector scalar tetraquark is predicted to be $m_{0^+}=3901.9\pm 12.5 \;\mathrm{MeV}$.

\appendix
\section{Explicit Expressions of the Spectra Densities}
\begin{align}
\rho_0(s) &=\frac{m_c^6}{{2}^{9} \pi ^6}{I^{}_{4,2,2,2}(s)} - \frac{m_c^8}{{2}^{9} {5} \pi ^6}{I^{}_{5,3,3,1}(s)}-\frac{7 {\langle g^2 G^2 \rangle} m_c^4}{{2}^{12} {3} \pi ^6}{I^{}_{3,2,2,1}(s)} + 
                   \frac{{\langle g^2 G^2 \rangle} m_c^2}{{2}^{11} \pi ^6}{I^{}_{2,1,1,2}(s)} + \nonumber \\ & 
                   \frac{{\langle g^2 G^2 \rangle} m_c^6}{{2}^{12} {3}^{2} \pi ^6}\left[9 {I^{\delta}_{2,2,2}(s)}+16 {I^{\delta}_{3,2,3}(s)}\right] - 
                   \frac{{\langle g^2 G^2 \rangle} m_c^8}{{2}^{12} {3} \pi ^6}{\left(\frac{\partial}{\partial s}I^{\delta}_{3,3,3}(s)\right)}-\nonumber \\ & 
                   \frac{{\langle \bar d d \rangle\langle \bar u u \rangle} m_c^6}{12 \pi ^2}{\left(\frac{\partial}{\partial s}I^{\delta}_{1,2,2}(s)\right)} -  
                   \frac{{\langle \bar d d \rangle\langle \bar u u \rangle} m_c^8}{{2}^{3} {3}^{2} \pi ^2}{\left(\frac{\partial^3}{\partial s^3}I^{\delta}_{3,3,3}(s)\right)} s+\nonumber \\ & 
                   \frac{{\langle \bar d d \rangle\langle \bar u u \rangle} m_0^2 m_c^6}{{2}^{3} {3} \pi ^2}{\left(\frac{\partial^3}{\partial s^3}I^{\delta}_{1,2,2}(s)\right)} s - \nonumber \\
                     & 
                   \frac{{\langle \bar d d \rangle\langle \bar u u \rangle} m_0^2 m_c^8}{{2}^{4} {3}^{2} \pi ^2}\left[2 {\left(\frac{\partial^3}{\partial s^3}I^{\delta}_{3,3,3}(s)\right)}+2
                     {\left(\frac{\partial^4}{\partial s^4}I^{\delta}_{3,3,3}(s)\right)} s-{\left(\frac{\partial^5}{\partial s^5}I^{\delta}_{3,3,3}(s)\right)} s^2\right]-\nonumber \\ & 
                   \frac{{\langle \bar d d \rangle\langle \bar u u \rangle} {\langle g^2 G^2 \rangle} m_c^4}{{2}^{4} {3}^{2} \pi ^2}{\left(\frac{\partial^3}{\partial
                     s^3}I^{\delta}_{1,2,2}(s)\right)} s + \nonumber \\ & 
                   \frac{{\langle \bar d d \rangle\langle \bar u u \rangle} {\langle g^2 G^2 \rangle} m_c^6}{{2}^{4} {3}^{3} \pi ^2}\left[2 {\left(\frac{\partial^3}{\partial
                     s^3}I^{\delta}_{2,2,3}(s)\right)}+2 {\left(\frac{\partial^4}{\partial s^4}I^{\delta}_{2,2,3}(s)\right)} s-{\left(\frac{\partial^5}{\partial s^5}I^{\delta}_{2,2,3}(s)\right)}
                     s^2\right] + \nonumber \\ & 
                   \frac{{\langle \bar d d \rangle\langle \bar u u \rangle} m_0^4 m_c^6}{{2}^{6} {3} \pi ^2}\left[2 {\left(\frac{\partial^3}{\partial s^3}I^{\delta}_{1,2,2}(s)\right)}+2
                     {\left(\frac{\partial^4}{\partial s^4}I^{\delta}_{1,2,2}(s)\right)} s-{\left(\frac{\partial^5}{\partial s^5}I^{\delta}_{1,2,2}(s)\right)} s^2\right] - \nonumber \\ & 
                   \frac{{\langle \bar d d \rangle\langle \bar u u \rangle} m_0^4 m_c^8}{{2}^{7} {3}^{2} \pi ^2}\left[12 {\left(\frac{\partial^4}{\partial s^4}I^{\delta}_{3,3,3}(s)\right)}-6
                     {\left(\frac{\partial^6}{\partial s^6}I^{\delta}_{3,3,3}(s)\right)} s^2+{\left(\frac{\partial^7}{\partial s^7}I^{\delta}_{3,3,3}(s)\right)} s^3\right]-\nonumber \\ & 
                   \frac{{\langle \bar d d \rangle\langle \bar u u \rangle} {\langle g^2 G^2 \rangle} m_0^2 m_c^4}{{2}^{5} {3}^{2} \pi ^2}\left[2 {\left(\frac{\partial^3}{\partial
                     s^3}I^{\delta}_{1,2,2}(s)\right)}+2 {\left(\frac{\partial^4}{\partial s^4}I^{\delta}_{1,2,2}(s)\right)} s-{\left(\frac{\partial^5}{\partial s^5}I^{\delta}_{1,2,2}(s)\right)}
                     s^2\right] + \nonumber \\ & 
                   \frac{{\langle \bar d d \rangle\langle \bar u u \rangle} {\langle g^2 G^2 \rangle} m_0^2 m_c^6}{{2}^{5} {3}^{3} \pi ^2}\left[12 {\left(\frac{\partial^4}{\partial
                     s^4}I^{\delta}_{2,2,3}(s)\right)}-6 {\left(\frac{\partial^6}{\partial s^6}I^{\delta}_{2,2,3}(s)\right)} s^2+{\left(\frac{\partial^7}{\partial s^7}I^{\delta}_{2,2,3}(s)\right)}
                     s^3\right]-\nonumber \\ & 
                   \frac{{\langle \bar d d \rangle\langle \bar u u \rangle} {\langle g^2 G^2 \rangle} m_0^4 m_c^4}{{2}^{8} {3}^{2} \pi ^2}\left[12 {\left(\frac{\partial^4}{\partial
                     s^4}I^{\delta}_{1,2,2}(s)\right)}-6 {\left(\frac{\partial^6}{\partial s^6}I^{\delta}_{1,2,2}(s)\right)} s^2+{\left(\frac{\partial^7}{\partial s^7}I^{\delta}_{1,2,2}(s)\right)}
                     s^3\right] + \nonumber \\ & 
                   \frac{{\langle \bar d d \rangle\langle \bar u u \rangle} {\langle g^2 G^2 \rangle} m_0^4 m_c^6}{{2}^{8} {3}^{3} \pi ^2}\left[72 {\left(\frac{\partial^5}{\partial
                     s^5}I^{\delta}_{2,2,3}(s)\right)}-48 {\left(\frac{\partial^6}{\partial s^6}I^{\delta}_{2,2,3}(s)\right)} s- 
                     \right. \nonumber \\& \left.  24 {\left(\frac{\partial^7}{\partial
                     s^7}I^{\delta}_{2,2,3}(s)\right)} s^2+12 {\left(\frac{\partial^8}{\partial s^8}I^{\delta}_{2,2,3}(s)\right)} s^3-{\left(\frac{\partial^9}{\partial
                     s^9}I^{\delta}_{2,2,3}(s)\right)} s^4\right] 
\end{align}
\begin{align}
    \rho_1(s) &= - \rho^{0^{+}}(s) +\frac{m_c^8}{{2}^{10} {15} \pi ^6}{I^{}_{6,3,3,1}(s)}+
                    \frac{{\langle g^2 G^2 \rangle} m_c^2}{{2}^{11} \pi ^6}{I^{}_{2,1,1,2}(s)} + 
                    \nonumber \\ & 
                    \frac{{\langle g^2 G^2 \rangle} m_c^4}{{2}^{14} {3}^{2} \pi ^6}\left[16 {I^{}_{3,2,2,1}(s)}-{I^{}_{4,2,2,1}(s)}\right] - 
                    \frac{{\langle g^2 G^2 \rangle} m_c^6}{{2}^{11} {3}^{2} \pi ^6}{I^{\delta}_{4,2,3}(s)} + 
                    \nonumber \\ & 
                    \frac{{\langle g^2 G^2 \rangle} m_c^8}{{2}^{14} {3} \pi ^6}{\left(\frac{\partial}{\partial s}I^{\delta}_{4,3,3}(s)\right)}-
                    \frac{{\langle \bar d d \rangle\langle \bar u u \rangle} m_c^8}{{2}^{3} {3}^{2} \pi ^2}{\left(\frac{\partial^2}{\partial
                      s^2}I^{\delta}_{3,3,3}(s)\right)}+
                    \nonumber \\ & 
                    \frac{{\langle \bar d d \rangle\langle \bar u u \rangle} m_0^2 m_c^8}{{2}^{4} {3}^{2} \pi ^2}{\left(\frac{\partial^4}{\partial
                      s^4}I^{\delta}_{3,3,3}(s)\right)} s-
                    \nonumber \\ & 
                    \frac{{\langle \bar d d \rangle\langle \bar u u \rangle} {\langle g^2 G^2 \rangle} m_c^6}{{2}^{5} {3}^{3} \pi ^2}{\left(\frac{\partial^4}{\partial
                      s^4}I^{\delta}_{2,2,3}(s)\right)} s + 
                    \nonumber \\ & 
                    \frac{{\langle \bar d d \rangle\langle \bar u u \rangle} m_0^4 m_c^8}{{2}^{7} {3}^{2} \pi ^2}\left[2 {\left(\frac{\partial^4}{\partial
                      s^4}I^{\delta}_{3,3,3}(s)\right)}+2 {\left(\frac{\partial^5}{\partial s^5}I^{\delta}_{3,3,3}(s)\right)} s-{\left(\frac{\partial^6}{\partial
                      s^6}I^{\delta}_{3,3,3}(s)\right)} s^2\right]-
                    \nonumber \\ & 
                    \frac{{\langle \bar d d \rangle\langle \bar u u \rangle} {\langle g^2 G^2 \rangle} m_0^2 m_c^6}{{2}^{6} {3}^{3} \pi ^2}\left[2
                      {\left(\frac{\partial^4}{\partial s^4}I^{\delta}_{2,2,3}(s)\right)}+2 {\left(\frac{\partial^5}{\partial s^5}I^{\delta}_{2,2,3}(s)\right)}
                      s-{\left(\frac{\partial^6}{\partial s^6}I^{\delta}_{2,2,3}(s)\right)} s^2\right]-
                    \nonumber \\ & 
                    \frac{{\langle \bar d d \rangle\langle \bar u u \rangle} {\langle g^2 G^2 \rangle} m_0^4 m_c^6}{{2}^{9} {3}^{3} \pi ^2}\left[12
                      {\left(\frac{\partial^5}{\partial s^5}I^{\delta}_{2,2,3}(s)\right)}-6 {\left(\frac{\partial^7}{\partial s^7}I^{\delta}_{2,2,3}(s)\right)}
                      s^2+{\left(\frac{\partial^8}{\partial s^8}I^{\delta}_{2,2,3}(s)\right)} s^3\right] 
\end{align}
\begin{align}
    \rho_T(s)&=2 \rho^{1^+}(s) +
                    \frac{{\langle g^2 G^2 \rangle} m_c^4}{{2}^{13} {3}^{2} \pi ^6}\left[4 {I^{}_{3,2,2,1}(s)}-{I^{}_{4,2,2,1}(s)}\right] - 
                    \nonumber \\ & 
                    \frac{{\langle g^2 G^2 \rangle} m_c^6}{{2}^{11} \pi ^6}{I^{\delta}_{2,2,2}(s)} + 
                    \nonumber \\ & 
                    \frac{{\langle g^2 G^2 \rangle} m_c^8}{{2}^{13} {3} \pi ^6}\left[4 {\left(\frac{\partial}{\partial s}I^{\delta}_{3,3,3}(s)\right)}+{\left(\frac{\partial}{\partial
                      s}I^{\delta}_{4,3,3}(s)\right)}\right]
\end{align}
where 
\begin{align}
    I^{\delta}_{n,m,l}(s) & \equiv \int_0^1 dx \int_0^{1-x} dy \frac{(1-x-y)^n}{x^m y^l} \delta(s-s(x,y)) \nonumber\\
    I^{}_{n,m,l,k}(s) & \equiv \frac{1}{\Gamma(k)} \int_0^1 dx \int_0^{1-x} dy \frac{(1-x-y)^n}{x^m y^l} (s-s(x,y))^{k-1}\theta(s-s(x,y)) \nonumber\\
    s(x,y) &\equiv \frac{m_c^2}{x} + \frac{m_c^2}{y}
\end{align}

\bibliographystyle{ieeetr}

\bibliography{makale}

\end{document}